\documentclass[11pt,amsmath,amssymb,nofootinbib]{revtex4}
\usepackage{amsmath}
\usepackage{amsfonts}
\usepackage{amssymb}
\usepackage[dvips]{graphicx}

\begin{document}

\newcommand\be{\begin{equation}}
\newcommand\ee{\end{equation}}
\newcommand\ba{\begin{eqnarray}}
\newcommand\ea{\end{eqnarray}}
\newcommand\bseq{\begin{subequations}} 
\newcommand\eseq{\end{subequations}}
\newcommand\bcas{\begin{cases}}
\newcommand\ecas{\end{cases}}
\newcommand{\p}{\partial}
\newcommand{\f}{\frac}
\newcommand{\nn}{\nonumber \\}
\def\tr{{\rm Tr}}

\title{A brief overview of quantum field theory with deformed symmetries and their relation with quantum gravity}

\author{Antonino Marcian\`o}

\address{Centre de Physique Th\'eorique,\\
Case 907 Luminy, 13288 Marseille, EU\\
E-mail: antonino.marciano@cpt.univ-mrs.fr}

\begin{abstract}

\noindent In this letter we outline some reasons for considering a quantum field theory symmetric under quantum groups and we sketch some results obtained with collaborators in the $\kappa$-Poincar\'e framework. We deal with this latter as a toy model towards an effective and low-energy theory of quantum gravity, the new physically relevant effects of which are Planck-scale suppressed.

\end{abstract}

\maketitle
\bigskip

\noindent 

The main reason for taking account of quantum field theory symmetric under a deformation of Poincar\'e algebra is the aim of introducing Planck-scale-suppressed effects. The relation between quantum groups and quantum gravity already emerges in some three-dimensional approaches to the full theory. Indeed, new physically relevant effects of quantum gravity are often thought to be related to the Planck scale (the Planck length being $L_p\simeq 1,6\cdot 10^{-34} \,m$), and in several models such an expected relation has been rigourously derived \cite{Rovelli}.

Quantum groups\cite{qgis123,qgis567} are (Hopf algebras) called ``quantum'' because they are obtained by a deformation of Poisson Lie algebras which shows similarities with the Moyal quantization of Poisson manifolds, leading from classical mechanics to non-relativistic quantum mechanics. From a purely algebraic point of view, the way in which new physics enter the scheme of effective theories enjoying such a symmetry structure is by the deformation of standard Lie algebra symmetries by means of a dimensionful parameter\cite{Majid:1988we, Lukierski:1992dt1} (proportional to $L_p$). It results a theory that has a relation to the standard Lie-algebra symmetric one which resembles the links one finds among the Galileian theory and the theory of special relativity. 

But this kind of ``quantized" symmetries also emerges in the low-energy limit of certain quantum gravity models. Indeed a quantum field theory enjoying quantum groups of symmetries has been derived from at least three different perspectives in $2+1$D quantum gravity.  In Ref.\cite{Schroers}, starting from the Chern-Simons formulation of $2+1$D gravity,  it has been shown that gravitational interactions deform the Poincar\'e symmetry of flat space-time to a quantum group symmetry, namely the Lorentz double. In particular, the Hilbert space of two gravitating particles has been there studied by means of the universal $R$-matrix of the Lorentz double.

In Ref.~\cite{Freidel&LivinePRL} the authors considered the issue of semi-classical regime in quantum gravity. They showed how to recover standard quantum field theory amplitudes in the no-gravity limit and how to compute the quantum gravity corrections, moving from Ponzano-Regge amplitudes with massive particles. Their conclusion has been that the effective dynamics of quantum particles coupled to quantum 3D gravity can be expressed in terms of an effective field theory symmetric under a quantum groups-like deformation of the Poincar\'e group.

Another reason for considering effective theories enjoying quantum groups (as an effective model for quantum gravity) relies on the fact that several approaches to quantum gravity have led to arguments in favour of the emergence of a minimum length and/or a maximum momentum, possibly related to the Planck length. For instance, concerning effective theories, this reason has led some authors---see, {\it e.g.}, Refs. \cite{Susskind, Garay:1994en}---to argue that somehow boosts should saturate at the Planck scale\cite{kowa}. A toy model for such an effective theory is represented by $\kappa$-Poincar\'e symmetric theories. 

An important property of the quantum groups symmetric picture is that it provides the best arena for implementing the ideas of deforming without breaking space-time Poincar\'e symmetries\cite{dsr1,dsr2}. Indeed the additional (planckian) parameter regulates quantum gravitational effects in the semi-classical regime of space-time\cite{kowa2}. Moreover, these quantum groups symmetric toy models give new insight for Planck-scale-suppressed physics. For instance, at the level of effective quantum field theory there is a deep relation between ``quantum groups" and Planck-scale-suppressed non locality (determined by dynamical quantum gravity). This has been first emphasized in Ref.~\cite{Arzano:2007nx} and its consequences at the level of Fock space quantization of scalar fields, deformed symmetries and entanglement have been analyzed in Ref.~\cite{ArMa2, Arzano-Alioscia}. Far from locality, symmetries of the theory cannot be described in terms of Lie algebras but need a new language, provided by non co-commutative Hopf algebras namely quantum groups.

While dealing with toy-models, we are equally concerned with various interpretational problems of different difficulty. The first one regards the nature of these nontrivial Hopf-algebra space-time symmetries and their characterization in terms of physical (conserved) quantities. This point of investigation stands as a primary one, although it has been consistently addressed only recently. 
Results reported in \cite{kappa1, kappa2,kappa3, theta, ArMa1} provided a first Noether-analysis of conserved quantities associated to $\kappa$-Poincar\'e\cite{Majid-Ruegg} and to $\theta$-Poincar\'e\cite{Chaichian:2004yh2} Hopf-algebras.
In both these frameworks  the necessity of an interplay between symmetry transformations has been pointed out in \cite{theta, kappa2}. Such results provided encouragement for the idea that these Hopf-algebra symmetries are truly meaningful in characterizing observable aspects of the relevant theories.
  
More clearness about this latter point encouraged a first insight in the construction of an effective quantum field theory playing as a toy-model endowed with all the ``good'' properties listed above. The most important feature we wanted hence to implement in the model\cite{ArMa2} has been the quantum groups symmetry consistency. In \cite{ArMa2} we proposed an approach to the quantization of $\kappa$-Poincar\'e symmetric fields, whose key-difference from previous studies is that Fock-space structure of the quantum states must be compatible with the algebraic and co-algebraic structure of the associated $\kappa$-Poincar\'e Hopf-algebra of symmetries. 
The two main features of this space of quantum states are the existence of a (symmetries-induced) planckian cut-off $\kappa$ for the field-modes in the one-particle sector and the need of a non-trivial symmetrization in the construction of the Fock space of the theory. It has been then found out that energy $Q_0$ and momentum $\vec{Q}$ charges of a single particle state obey to the $\kappa$-deformed dispersion relation
\begin{center}
$|\vec{Q}|=\kappa \tanh\left(\frac{Q_0}{\kappa}\right)\, .$\\
\end{center}
This type of non-trivial $\kappa$-deformed-Fock-space, characterized by a deformation of (multiparticle modes') symmetrization which is $\kappa$-suppressed, radically differs from the more popular framework of ``twisted'' (\i.e. $\theta$-deformed) statistics (see for instance references in Ref~\cite{MarcianoArab}). The features of such a non-trivial symmetrization can indeed introduce entanglement\cite{Arzano-Alioscia} at the Planck scale for multiparticle states.
  
The work done until now is still preliminary and we are far from claiming to have a sufficiently developed candidate for an effective theory enjoying Planck-scale-suppressed effects of physical relevance. Nevertheless richness of  the mathematical structure (implementing a huge variety of new ideas in physics), existence of derivations from the full theory of quantum gravity and last but not least suitability of phenomenological applications  are strongly encouraging the quantum gravity community to work towards this direction.

\end{document}